\documentclass[12pt]{article}

\usepackage[cp1251]{inputenc}
\usepackage{graphicx}
\usepackage{epstopdf}
\usepackage{amssymb}
\usepackage{amsmath}
\usepackage{wrapfig}
\usepackage{hyperref}

\newcommand{\rarr}{\rightarrow}

\newcommand{\pd}{\partial}
\newcommand{\la}{\langle}
\newcommand{\ra}{\rangle}
\newcommand{\bs}{\boldsymbol}
\newcommand{\ts}{\textstyle}
\newcommand{\ol}{\overline}
\newcommand{\const}{\mathop{\rm const}\nolimits}
\newcommand{\sign}{\mathop{\rm sign}\nolimits}

\begin{document}


\thispagestyle{empty}

\begin{titlepage}


\vspace{0.3cm}

\begin{center}
    \Large \bf
    Synchrotron Radiation in the Standard Model Extension
\end{center}


\vspace{0.5cm}

\renewcommand{\thefootnote}{\fnsymbol{footnote}}

\begin{center}
    I.~E.~Frolov\footnotemark[1], V.~Ch.~Zhukovsky\footnotemark[2], \\

    {
        \sl
        Faculty of Physics,
        Department of Theoretical Physics, \\ 
        Moscow State University,
        119899, Moscow, Russia.
    }
\end{center}

\vspace{0.6cm}


\begin{abstract}
We obtain a system of exact solutions of the Dirac equation for an electron moving in a constant homogeneous external magnetic field with account of its vacuum magnetic moment and assumed Lorentz invariance violation in the minimal CPT-odd form in the framework of the Standard Model Extension. Using these solutions, 
characteristics of the particle synchrotron radiation are calculated, and possible observable effects caused by the Lorentz non-invariant interaction are described. We demonstrate that the angular distribution of the radiation has specific asymmetry, which can be explained as a consequence of non-conservation of transversal electron polarization in the presence of a background Lorentz non-invariant condensate field.
\end{abstract}


\vspace{0.3cm}

\setcounter{footnote}{1}
\noindent
\footnotetext{\noindent frolov\underline{\ }ie@mail.ru} 

\setcounter{footnote}{2}
\noindent
\footnotetext{\noindent zhukovsk@phys.msu.ru}

\vfill
\setcounter{footnote}{0}


\end{titlepage}


\section{Introduction}
\label{intro}

According to the modern viewpoint, 
the standard quantum field theory model of elementary particles is a low-energy approximation of a certain more fundamental theory 
which unites in some way all known types of physical interactions together. As a consequence there must exist (and, in spite of their extreme subtleness, be observable) specific effects, which are not inherent in the standard model and which reveal some features of the underlying more profound theory. In particular, the Lorentz and CPT-symmetry breaking may be expected to occur for physical particles, caused by some dynamic reasons lying outside the standard model. The theoretical framework, which covers the standard model and includes phenomenological description of Lorentz invariance violation in a rather general form, is known as the Standard Model Extension (SME)
    \cite{sme_theory}.

    The emergence of the SME yielded intense theoretical investigation and discussion of its implications for particle physics \cite{sme_calc}, \cite{sme_reviews}. Among new phenomena predicted and already partially available for high-precision observations, one can mention the following: ``distortion'' of energy-momentum relations of physical particles, effective anisotropy and dispersion of the vacuum (due to the nature of the SME, this effect is inevitably present in almost all situations considered);
non-trivial differences in properties of particles and antiparticles, and
particles with different helicities, additional anomalous magnetic moments
    \cite{Andrianov_Giacconi_Soldati}, \cite{sme_eff_particles}, \cite{sme_vmm_theory}, \cite{sme_vmm_from_phot};
possible new channels of high-energy reactions (for instance, processes like $ e^{-}
\rarr e^{-} + \gamma $, $ \gamma \rarr e^{+} + e^{-} $ \cite{Zhukovsky_Lobanov_Murchikova});
specific asymmetry of angular distributions of radiation and reaction products (an example of this
kind is shown in the present work); effects at finite temperature \cite{Ebert_Zhukovsky_Razumovsky}; various
    astrophysical manifestations \cite{Andrianov_Giacconi_Soldati}, \cite{sme_eff_astro}, \cite{Altschul}; etc.

    In the present work, we shall actually consider a very particular limit of the SME related to the minimal Lorentz and CPT-symmetry breaking in the electron sector of the theory (the reasons for making this approximation are discussed further in the paper in more detail);
it is described by means of the additional (to the standard fermion Lagrangian) term $ -\ol{\psi} \gamma^{5} \gamma^{\mu} b_{\mu} \psi $, where $ b^{\mu} $ is a pseudovector
quantity independent of the
space-time coordinates $ x^{\mu} $ in every fixed reference frame, regarded as a vacuum expectation value of some object of a more fundamental theory.

We shall study the synchrotron radiation of a high-energy electron supposing (in the same way as it was 
done in \cite{Zhukovsky_Lobanov_Murchikova}) that $ | b^{0} | \gg | {\bf b} | $; such assumption does not contradict most of the
estimates available at present (see, e.g., \cite{sme_reviews}, \cite{Andrianov_Giacconi_Soldati}). So 
we consider the interaction of the fermion field $ \psi $ with the condensate $ b^{\mu} $ as the only type of Lorentz invariance violation present in our theory, neglecting other possible types (in particular, referring to the photon sector).

    In contrast to the earlier publications on the similar subject, where only classical approach was adopted (see, e.g., \cite{Altschul}, and also \cite{Montemayor_Urrutia}, \cite{Ellis_Mavromatos_Sakharov}), we shall carry out consistent quantum consideration of the phenomena, including {\it both} the motion of an electron {\it and} its radiation.
We take Lorentz invariance violation into account exactly, i.e. avoiding (at the stage of quantization) the perturbation theory approximation and considering $ b^{0} $ as a given classical external field in addition to the magnetic field background. We obtain a system of exact solutions of the modified Dirac equation for an electron with a vacuum magnetic moment in a constant magnetic field, which is necessary for the quantum description of the fermion field in the Furry picture \cite{Furry}.
    Employing the methods of QED, with the use of these solutions, we calculate the asymptotic expressions for the spectral-angular distribution of the electromagnetic radiation in the weak-field limit which incorporate yet {\it all} the quantum corrections arising, and we pay special attention to the radiation properties caused by the Lorentz invariance violation assumed in the theory.

    Throughout the work, the natural units $ \hbar = c = 1 $ are adopted.


\section{Motivations} 
\label{theory}

In this section, let us recall some of the SME principles and consider the issues related to a high-energy electron handled within this framework. The Lagrangian of the SME is obtained as the most general one describing the possible Lorentz non-invariant interactions and consistent with the fundamental principles of the modern quantum field theory (such as gauge invariance and power-counting renormalizability) \cite{sme_theory}. It is believed that these interactions are due to the spontaneous Lorentz symmetry breaking in the underlying theory, and this imposes the requirement that the corresponding Lagrangian terms are Lorentz scalars. Consequently, in the low-energy limit, there emerges the extended quantum electrodynamics from the lepton-gauge sector of the SME. Its Lagrangian can be written in the following general form:
\begin{equation}
    \label{theory_L}
    {\cal L}^\text{QED Extension} =
        {\cal L}^\text{QED}
        + {\cal L}_\text{lepton}
        + {\cal L}_\text{photon},
\end{equation}
where $ {\cal L}^\text{QED} $ is the well-known QED Lagrangian:
\begin{equation}
    {\cal L}^\text{QED} =
        - \frac{1}{4} F_{\mu \nu} F^{\mu \nu}
        + \frac{1}{2} \overline{\psi} i \gamma^{\mu} \overset{\leftrightarrow}{D_{\mu}} \psi
        - \overline{\psi} m \psi,
\end{equation}
$ {\cal L}_\text{lepton} $, $ {\cal L}_\text{photon} $ are the terms describing the Lorentz non-invariant interactions in the electron and the photon sector of the theory respectively. They can be decomposed into the sum of the CPT-even and odd parts:
\begin{subequations}
\begin{equation}
    \label{theory_L_even_lepton}
    {\cal L}^{\rm CPT-even}_{\rm lepton} =
        - \frac{1}{2} \overline{\psi} \sigma^{\mu \nu} H_{\mu \nu} \psi
        + \frac{1}{2} \overline{\psi}
            i \gamma^{\mu} c_{\mu \nu} \overset{\leftrightarrow}{D^{\nu}} \psi
        + \frac{1}{2} \overline{\psi}
            \gamma^{5} i \gamma^{\mu} d_{\mu \nu} \overset{\leftrightarrow}{D^{\nu}} \psi,
\end{equation}
\begin{equation}
    \label{theory_L_odd_lepton}
    {\cal L}^{\rm CPT-odd}_{\rm lepton} =
        - \overline{\psi} \gamma^{\mu} a_{\mu} \psi
        - \overline{\psi} \gamma^{5} \gamma^{\mu} b_{\mu} \psi,
\end{equation}
\begin{equation}
    \label{theory_L_even_photon}
    {\cal L}^{\rm CPT-even}_{\rm photon} =
        -\frac{1}{4} (k_{F})_{\alpha \beta \mu \nu} F^{\alpha \beta} F^{\mu \nu},
\end{equation}
\begin{equation}
    \label{theory_L_odd_photon}
    {\cal L}^{\rm CPT-odd}_{\rm photon} =
        \frac{1}{2} (k_{AF})^{\alpha} \varepsilon_{\alpha \beta \mu \nu} A^{\beta} F^{\mu \nu}.
\end{equation}
\end{subequations}
The dimensionless coefficients $ c_{\mu \nu} $, $ d_{\mu \nu} $, $ (k_{F})_{\alpha \beta \mu \nu} $ and the coefficients $ H_{\mu \nu} $, $ a_{\mu} $, $ b_{\mu} $, $ (k_{AF})_{\alpha} $ with dimensions of mass are the tensor parameters independent of the space-time coordinates $ x^{\mu} $, encoding the information relevant to the Lorentz invariance violation for physical particles in any given reference frame (for further details see \cite{sme_theory}, \cite{sme_reviews} and other reviews of the SME).

At present, most of the coefficients describing Lorentz invariance violation in the extended QED have been tightly constrained from the existing experimental data, including that of the experiments aimed at the search of the essentially new phenomena mentioned in the Introduction \cite{sme_reviews}. The bounds available have different order of magnitude for different sets and combinations of the coefficients (it is implied, of course, that the quantities with the same dimensionality are being compared), and it turns out that there still exists at least one relatively loosely constrained parameter of dimensions of mass. Considering the CPT-odd fermion Lagrangian term $ -\ol{\psi} \gamma^{5} \gamma^{\mu} b_{\mu} \psi $, one may argue \cite{Andrianov_Giacconi_Soldati} that the upper bound on the timelike component of $ b_{\mu} $ may be $ | b^{0} | < 10^{-2} \, \text{eV} $, while for the spacelike components one has $ | {\bf b} | < 10^{-18} \, \text{eV} $ or even stronger, and the remaining tensor coefficients are also estimated to be smaller than the upper bound on $ b^{0} $ \cite{sme_reviews}. This makes it reasonable to study the effects caused by the non-zero parameter $ b^{0} $ in the first place since it could actually be the greatest one among the others (with the same dimensions). One of the aims of our investigation is to examine the consequences of this situation for the synchrotron radiation and to place bounds on $ b^{0} $ possibly more stringent than those available in the literature.

At the same time, one may argue that the dimensionless coefficients (that are present in the kinetic-like terms of (\ref{theory_L_even_lepton})--(\ref{theory_L_odd_photon})) do not actually play a dominant role when considering any high-energy processes unless going into the Planck-scale physics (this differs significantly from the viewpoint accepted in \cite{Altschul}). In fact, it is believed that the scale that governs the values of the SME-specific parameters is the Planck scale, namely, the Planck mass $ M_{P} \simeq 1.22 \cdot 10^{28} \, \text{eV} $. This implies that coefficients with higher mass dimensions have to possess additional factors of $ M_{P} $ with respect to the others \cite{sme_theory}. It is obvious then that the Lagrangian terms containing the coefficients with the highest possible mass dimensions (i.e., $ mass^{1} $) are of primary importance for a particle with a given energy $ E \ll M_{P} $ since they do not involve any derivatives. In particular, if one has the relation $ | b_{\mu} | \sim M_{P} \, | c_{\mu \nu} | $ for the characteristic scales of the parameters $ b_{\mu} $ and $ c_{\mu \nu} $ then the average values of the corresponding Lagrangian terms are related as follows: $ | \la i \gamma^{\mu} c_{\mu \nu} D^{\nu} \ra | \sim \frac{E}{M_{P}} \, | \la -\gamma^{5} \gamma^{\mu} b_{\mu} \ra | $, and this is also valid for the other appropriate terms in (\ref{theory_L_even_lepton})--(\ref{theory_L_odd_photon}) as well. Thus, one may neglect the effects caused by non-zero dimensionless Lorentz-violating coefficients in the extended QED when studying the dominant phenomena at feasible energies.

Taken together, the arguments presented above justify the choice of a simplified model (obtained from the extended QED) in which there exists only one type of Lorentz invariance violation described by means of the CPT-odd fermion Lagrangian term $ -\ol{\psi} \gamma^{5} \gamma^{\mu} b_{\mu} \psi $ where $ b^{\mu} = \{ b, {\bf 0} \} $. As we believe, this is appropriate for investigating whether $ | b^{0} | $ is indeed much greater than other related parameters of the extended QED by way of studying the characteristics of the synchrotron radiation.

It should be noted, however, that there is an important phenomenon that cannot be neglected in our case when considering the motion of an electron in a magnetic field: Vacuum magnetic moment of the particle. Indeed, provided $ \mu \simeq \mu_{0} \, \frac{e^{2}}{2\pi} $ where $ \mu_{0} = \frac{e}{2m} $ \cite{Schwinger_vmm}, the characteristic energy of the corresponding interaction is $ \mu H \sim 10^{-6} \, \text{eV} $ in the typical laboratory field $ H \sim 10^{4} \, \text{gauss} $, so it may be of the same order of magnitude as the quantity $ b^{0} $ (or may be less or greater, the latter being somewhat more likely). Thus, we may not neglect the influence of the vacuum magnetic moment when studying the synchrotron radiation with account of the Lorentz invariance violation in the form we have chosen, and it turns out that the electron vacuum magnetic moment does interact with the condensate $ b_{\mu} $ in a nontrivial way giving rise to the prevailing observable effects (see the main text below).

We have considered the Lagrangian (\ref{theory_L}) in the context of the SME which is based on the conjecture of spontaneous Lorentz symmetry breaking in a more fundamental and profound theory, e.g., string theory (this is the approach generally adopted in \cite{sme_theory}). At the same time, similar and even the same terms as those present in (\ref{theory_L}) can arise under some other circumstances. In particular, it has been shown \cite{Kostelecky_liv_grav} that certain weak background effects possible in some generalized theories of gravity should also affect the SME-specific coefficients in (\ref{theory_L_even_lepton})--(\ref{theory_L_odd_photon}), and concerning the quantity $ b_{\mu} $, one can obtain the following leading-order expression for its effective value:
\begin{equation}
    \left( b_{\rm eff} \right)_{\mu} =
        b_{\mu} -
        {\ts \frac{1}{4}} \pd^{\alpha} \chi^{\beta \gamma} \epsilon_{\alpha \beta \gamma \mu} +
        {\ts \frac{1}{8}} T^{\alpha \beta \gamma} \epsilon_{\alpha \beta \gamma \mu}.
\end{equation}
Here $ \chi_{\mu a} $ is the antisymmetric part of the vierbein fluctuation against the Minkowski space-time background and $ T^{\lambda}{}_{\mu \nu}$ is the torsion tensor of the Riemann-Cartan space-time (for more detail, see \cite{Kostelecky_liv_grav} and references therein). Thus, global space-time curvature and possible torsion of the Universe obviously lead to the effects similar to those of the spontaneous Lorentz invariance breaking, so that they may interfere in a peculiar way enhancing or cancelling each other (see also \cite{Shapiro_liv_grav}). Moreover, a mechanism generating similar effects of Lorentz invariance violation involving chiral fermions has also been proposed \cite{liv_from_chiral_fermions}. Anyway, in general, since the Lagrangian (\ref{theory_L}) and the experimental bounds on the SME-specific coefficients of the extended QED are actually model-independent (they are derived using the standard quantum field theory techniques), the origin of these coefficients does not affect the way we are going to investigate the model and examine the phenomena arising, provided that the effective Lorentz-violating parameters remain constant in the space-time region being large enough (and this is usually assumed when working in the context of the SME).

In what follows, we study the motion of an electron in an external magnetic field at the quantum level and obtain the characteristics of its electromagnetic radiation using the techniques of QED.


\section{The model}
\label{model}

    The Lagrangian obtained
for an electron interacting with the electromagnetic field $ A^{\mu} $ and background constant condensate field $ b^{\mu} $ has the form:
\begin{equation}
    \label{model_L}
    {\cal L} =
        \ol{\psi} \left(
            i \gamma^{\alpha} D_{\alpha} - m
            + {\ts \frac{1}{2}} \mu \sigma^{\alpha \beta} F_{\alpha \beta}
            - \gamma^{5} \gamma^{\alpha} b_{\alpha}
        \right) \psi.
\end{equation}
Here $ F_{\alpha \beta} $ is the electromagnetic field tensor; $ \mu $ is the electron anomalous (vacuum) magnetic moment \cite{Pauli_vmm}, which we treat approximately as a constant quantity: $ \mu \simeq \mu_{0} \, \frac{e^{2}}{2\pi} $
where $ \mu_{0} = \frac{e}{2m} $ \cite{Schwinger_vmm}; $ \sigma^{\mu \nu} = \frac{i}{2} [ \gamma^{\mu}, \gamma^{\nu} ] $, $ \gamma^{5} = - i \gamma^{0} \gamma^{1} \gamma^{2} \gamma^{3} $. We accept $ e > 0 $
so that the electron charge $ q_{e} = -e $, and the covariant derivative is $ D_{\mu} = \pd_{\mu} - i e A_{\mu} $. We assume that in the laboratory reference frame $ b^{\mu} = \{ b, {\bf 0} \} $, $ b = \const $; let there also exist a constant homogeneous external magnetic field oriented along the $ z $-axis: $ {\bf H} = H {\bf e}_{z} $, $ H > 0 $.

One can, considering an electron with a definite energy so that $ \psi(t,{\bf r}) = e^{-iEt} \,
\Psi({\bf r}) $, represent the equations of motion for the field $ \psi $ resulting from (\ref{model_L}) in the hamiltonian form:
\begin{equation}
    \label{sol_HPsi_EPsi}
    H_{\rm D} \Psi = E \Psi,
\end{equation}
where $ H_{\rm D} $ is the hermitian energy operator:
\begin{equation}
    \label{model_H}
    H_{\rm D} =
        \bs \alpha {\bf P} + \gamma^{0} m - e A^{0}
        + \mu H \gamma^{0} \Sigma_{3} - b \gamma^{5},
\end{equation}
$ {\bf P} = {\bf p} + e {\bf A} $ is the canonical quantum-mechanical momentum, $ {\bf p} = -i \nabla $; $ \bs \alpha = \gamma^{0} \bs \gamma $, $ \Sigma_{i} = \frac{1}{2} \epsilon_{ijk} \sigma^{jk} $.
In order to perform the quantization of the theory, 
one must solve the eigenvalue problem (\ref{sol_HPsi_EPsi}) and find a complete
system of the electron wave functions $ \{ \Psi \} $.


\section{Solution of the equations of motion}
\label{sol}

Let us take the electromagnetic potential of the external magnetic field in the axial-symmetric
form:
\begin{equation}
    A^{\mu} = \{ 0, {\bf A} \}, \qquad
    {\bf A} = {\ts \frac{1}{2}} \{ -H y, H x, 0 \}.
\end{equation}
It is obvious that $ \left[ p_{z}, H_{\rm D} \right] = 0 $, therefore we shall further consider the problem (\ref{sol_HPsi_EPsi}) on the subspace with a definite fixed $ p_{z} \equiv p \, $:
\begin{equation}
    \label{sol_Psi_phi}
    \Psi(x,y,z) =
        {\ts \frac{1}{\sqrt{2\pi}}} \, e^{i p z} \phi(x,y),
\end{equation}
so that (\ref{sol_HPsi_EPsi}) turns into
\begin{equation}
    \label{sol_Hphi_Ephi}
    H_{\rm D} \phi = E \phi,
\end{equation}
where in the expression (\ref{model_H}) for the $ H_{\rm D} $ one must take $ P_{3} = p $ so that $ {\bf P} = \{ \hat{P}_{1}, \hat{P}_{2}, p \} $ (we use the ``hat'' symbol to denote operators in order to distinguish them from $ c $-values where a confusion is possible). Let us introduce now
the ``mixing angle'' $ \delta $ as follows:
\begin{equation}
    \label{sol_vartheta_def}
    \delta = \arctan \frac{b}{\mu H},
\end{equation}
so that
\begin{equation}
    \begin{aligned}
        \mu H &= \tilde{\mu} H \cos \delta, \\
        b     &= \tilde{\mu} H \sin \delta, \\
    \end{aligned}
\end{equation}
where
\begin{equation}
    \label{sol_mueff_def}
    \tilde{\mu} H =
        \sqrt{ (\mu H)^{2} + b^{2} }.
\end{equation}
We shall name the quantity $ \tilde{\mu} $ an effective anomalous magnetic moment. Using the angle $ \delta $, we go over to the effective mass and momentum:
\begin{equation}
    \label{sol_mpeff_def}
    \begin{pmatrix}
        \tilde{m} \\
        \tilde{p} \\
    \end{pmatrix}
    =
    \begin{pmatrix}
        \cos \delta  & \sin \delta \\
        -\sin \delta & \cos \delta \\
    \end{pmatrix}
    \begin{pmatrix}
        m \\
        p \\
    \end{pmatrix}.
\end{equation}
It should be noted that the effective mass $ \tilde{m} $ may take negative values, when $ p < - m
\cot \delta $. It is easy to see that with the help of the unitary transformation
\begin{equation}
    U^{-1} H_{\rm D} U = \tilde{H}_{\rm D}, \qquad
\end{equation}
where
\begin{equation}
    \label{sol_U_def}
    U
    = \exp({\ts -\frac{\delta}{2} \gamma^{3}})
    = \cos{\ts \frac{\delta}{2}} - \gamma^{3} \sin{\ts \frac{\delta}{2}},
\end{equation}
the hamiltonian (\ref{model_H}) can be brought to the following form:
\begin{equation}
    \label{sol_tilde_H}
    \tilde{H}_{\rm D} =
        \bs \alpha \tilde{\bf P} + \gamma^{0} \tilde{m}
        + \tilde{\mu} H \gamma^{0} \Sigma_{3},
\end{equation}
where $ \tilde{\bf P} = \{ \hat{P}_{1}, \hat{P}_{2}, \tilde{p} \} $ ($ \hat{P}_{1} $, $ \hat{P}_{2} $ are the same as in the hamiltonian (\ref{model_H})). Thus the problem (\ref{sol_Hphi_Ephi}) is
formally equivalent to the problem
\begin{equation}
    \label{sol_tilde_Hphi_Ephi}
    \tilde{H}_{\rm D} \tilde{\phi} = E \tilde{\phi},
\end{equation}
since the operators $ \tilde{H}_{\rm D} $ and $ H_{\rm D} $ have identical eigenvalues, and their
eigenvectors are related by the transformation (\ref{sol_U_def}): $ \phi = U \tilde{\phi} $. The hamiltonian (\ref{sol_tilde_H}) formally describes an electron with an anomalous magnetic moment in an external magnetic field without Lorentz symmetry breaking. The corresponding problem (\ref{sol_tilde_Hphi_Ephi}) has been studied and solved in \cite{Ternov_Bagrov_Zhukovsky} (this
work, however, deals with the physical electron with mass greater than zero; nevertheless it can be easily seen that in the case $ \tilde{m} < 0 $ one can, with the help of one more unitary
transformation, $ \tilde{H}'_{\rm D} = \gamma^{5} \tilde{H}_{\rm D} \gamma^{5} $, effectively
perform the change in the hamiltonian: $ \tilde{m} \rightarrow -\tilde{m} $, $ \tilde{\mu} \rightarrow -\tilde{\mu} $; after that all results obtained in \cite{Ternov_Bagrov_Zhukovsky} can
be applied to $ \tilde{H}'_{\rm D} $).

Let us give here only the final results of the solution of the problem (\ref{sol_Hphi_Ephi}). The
energy values under investigation can be written as follows:
\begin{equation}
    \label{sol_E}
    E = \epsilon \sqrt{
        \left( \Pi + \tilde{\mu} H \right)^{2} + \tilde{p}^{2}
    }, \qquad
    \epsilon = \pm 1,
\end{equation}
where
\begin{equation}
    \label{sol_Pi_values}
    \Pi =
        \zeta \sqrt{ \tilde{m}^{2} + 2eHn }, \quad
    n = 0, 1, \ldots, \quad
    \zeta = \begin{cases}
        \pm 1, & n > 0, \\
        -\sign \tilde{m}, & n = 0. \\
    \end{cases}
\end{equation}
The quantity $ \Pi $ is the eigenvalue of the electron polarization (or spin) operator
\begin{equation}
    \label{sol_Pi_def}
    \hat{\Pi} =
        \Pi_{\bot} \cos \delta + \Pi_{\|} \sin \delta,
\end{equation}
where
\begin{equation}
    \Pi_{\bot} = m {\bf \Sigma} + i \gamma^{0} \gamma^{5} [{\bf \Sigma} \times {\bf P}], \qquad
    \Pi_{\|} = {\bf \Sigma} {\bf P},
\end{equation}
which can be diagonalized together with $ H_{\rm D} $; it is defined in an unambiguous way (when $
\tilde{\mu} \neq 0 $). This operator can be named a ``mixed'' spin operator since it is a
superposition of the well-known transversal $ \Pi_{\bot} $ and longitudinal $ \Pi_{\|} $ polarization parts with the coefficients $ \cos \delta $ and $ \sin \delta $ respectively (for the
properties of the electron spin operators see, e.g., \cite{book_rel_el}, \cite{book_hep_rad}). In
the expression (\ref{sol_Pi_values}) $ n $ is the principal quantum number; it can be easily proved that the corresponding integral of motion is $ {\left( \alpha_{1} P_{1} + \alpha_{2} P_{2}
\right)}^{2} $ with the eigenvalues $ 2eHn $. In the case of $ n = 0 $, the sign of $ \Pi $ (i.e.,
the spin orientation) as it follows from (\ref{sol_Pi_values}) takes a definite value. When $
\delta \neq 0 $ it depends on $ p $ (through $ \tilde{m} $, according to (\ref{sol_mpeff_def})).
Besides, it is clear that the form (\ref{sol_Pi_def}) is valid not only under the assumption
(\ref{sol_Psi_phi}) but also in the general case when $ P_{3} = p_{z} \equiv -i \pd_{z} $; moreover $ \Pi $ is a gauge-invariant quantity (as it contains only the gauge-invariant canonical momentum $ {\bf P} $).

The wave functions corresponding to the spectrum (\ref{sol_E}) in the polar coordinate system $ (r, \varphi) $ are as follows:
\begin{equation}
    \label{sol_phi}
    \phi(r,\varphi) =
        {\ts \frac{1}{\sqrt{2\pi}}} \, e^{i (n - s - 1/2) \varphi}
        \begin{pmatrix}
            c_{1}   \, e^{-i \varphi / 2} \, I_{n-1,s}(\rho) \\
            i c_{2} \, e^{ i \varphi / 2} \, I_{n,s}  (\rho) \\
            c_{3}   \, e^{-i \varphi / 2} \, I_{n-1,s}(\rho) \\
            i c_{4} \, e^{ i \varphi / 2} \, I_{n,s}  (\rho) \\
        \end{pmatrix}
        \sqrt{eH}, \quad
    \rho = {\ts \frac{1}{2}} eHr^{2},
\end{equation}
where $ I_{n,s}(\rho) $ are the Laguerre functions:
\begin{equation}
    I_{n,s}(\rho) =
        {\ts \sqrt{\frac{s!}{n!}}} \, e^{-\rho / 2} \rho^{(n - s)/2} L_{s}^{n-s}(\rho),
\end{equation}
expressed by means of the generalized Laguerre polynomials $ L^{l}_{s}(\rho) $:
\begin{equation}
    L^{l}_{s}(\rho) =
        {\ts \frac{1}{s!}} \, e^{\rho} \rho^{-l}
        {\ts \frac{d^{s}}{d\rho^{s}}} \left( e^{-\rho} \rho^{s + l} \right);
\end{equation}
$ s = 0, 1, \ldots $ is the radial quantum number; $ \{ c_{\mathfrak a} \} $ are constant coefficients depending on the particle state. The solutions (\ref{sol_phi}) are chosen to be the eigenfunctions of the $ z $-component of the fermion particle angular momentum operator $ J_{z} =
-i \frac{\pd}{\pd \varphi} + \frac{1}{2} \Sigma_{3} $, which corresponds to the axial symmetry
existing in our problem:
\begin{equation}
    J_{z} \psi = \left( l - {\ts \frac{1}{2}} \right) \psi, \qquad
    l = n - s.
\end{equation}

In the standard (or Dirac) representation of the $ \gamma $-matrices the coefficients $ \{
c_{\mathfrak a} \} $, which meet the normalization requirement for the wave functions
\begin{equation}
    \int r dr \, d\varphi \, \phi^{\dag} \phi^{ } = 1, 
\end{equation}
can be written as follows:
\begin{equation}
    \label{sol_coeffs}
    \setlength\arraycolsep{1pt}
    \begin{pmatrix}
        c_{1} \\
        c_{2} \\
        c_{3} \\
        c_{4} \\
    \end{pmatrix}
    = \frac{1}{2 \sqrt{2}}
    \begin{pmatrix}
        A ( P \alpha + \epsilon \zeta Q \beta )        \\
        -\zeta B ( P \alpha - \epsilon \zeta Q \beta ) \\
        A ( P \beta - \epsilon \zeta Q \alpha )        \\
        \zeta B ( P \beta + \epsilon \zeta Q \alpha )  \\
    \end{pmatrix},
\end{equation}
where
\begin{equation}
    \label{sol_ABPQ}
    \begin{aligned}
        A &= \sqrt{1 + \frac{\tilde{m}}{\Pi}}, & \qquad &
        P &= \sqrt{1 + \frac{\tilde{p}}{E}}, \\
        B &= \sqrt{1 - \frac{\tilde{m}}{\Pi}}, & \qquad &
        Q &= \sqrt{1 - \frac{\tilde{p}}{E}}, \\
    \end{aligned}
\end{equation}
and
\begin{equation}
    \label{sol_alpha_beta}
    \alpha = \cos{\ts \frac{\delta}{2}} - \sin{\ts \frac{\delta}{2}}, \qquad
    \beta  = \cos{\ts \frac{\delta}{2}} + \sin{\ts \frac{\delta}{2}}.
\end{equation}
The expression (\ref{sol_coeffs}) is valid for all $ n $ and $ \tilde{m} $. It should be noted that expressions (\ref{sol_E}), (\ref{sol_Pi_values}) for the energy $ E $ and the quantity $ \Pi $ and
expression (\ref{sol_coeffs}) for the coefficients $ \{ c_{\mathfrak a} \} $ do not include the
quantum number $ s $. This degeneracy is typical of the electron in a uniform magnetic field problem and is related to the invariance with respect to the choice of the position of the electron center of orbit.

Thus we have found the eigenvalues and obtained the orthonormalized eigenfunction system of the
hamiltonian $ H_{\rm D} $; the full set of quantum numbers is $ \{ n, s, p, \zeta, \epsilon \} $,
where
\begin{equation}
    n, s = 0, 1, \ldots, \quad
    -\infty < p < +\infty, \quad
    \zeta = \pm 1, \quad
    \epsilon = \pm 1.
\end{equation}
The wave functions and the energy spectrum are formally similar in structure to those of the problem without Lorentz invariance breaking considered in \cite{Ternov_Bagrov_Zhukovsky}. However in our case, the parameters $ \tilde{m} $, $ \tilde{p} $, $ \tilde{\mu} $ are effective quantities depending, as well as the coefficients (\ref{sol_coeffs}), on the mixing angle $ \delta $.


\section{Synchrotron radiation} 
\label{rad}

In this section, we investigate the electromagnetic radiation of an electron moving in a magnetic field using the wave functions obtained earlier. The radiation is handled at the entirely quantum level, in contrast to the classical approach adopted in most of the present-day publications on the similar subject (see, e.g., \cite{Altschul}, \cite{Montemayor_Urrutia}, \cite{Ellis_Mavromatos_Sakharov}). We shall calculate the asymptotic expressions for the spectral-angular distribution of the one-photon radiation of a high-energy electron in the weak-field limit. Unlike Schwinger's method that takes only the first quantum corrections into account \cite{Schwinger_sr}, we use the technique that provides us with {\it all} the quantum corrections arising. In this respect, our results are {\it exact} (i.e., we do not actually make any expansion in powers of $ \hbar $), although they are {\it asymptotic}, appropriate for the case of the weak magnetic field $ H \ll H_{c} $ (with respect to Schwinger's critical field $ H_{c} \simeq 4.41 \cdot 10^{13} \, \text{gauss} $), with the small parameter being the ratio of the mass of the electron to its energy.

It is essential to consider the radiation phenomena at the quantum level, since the effects we are interested in (related to the Lorentz invariance violation present in the theory) are actually due to the change of the spin state of the electron (described by coefficients (\ref{sol_coeffs}) of the wave functions (\ref{sol_phi})), and spin effects should disappear in the classical limit.


\subsection{General theory}

Consider the electron\footnote{We assume $ \epsilon = \epsilon' = +1 $ throughout this section; the generalization for positrons is straightforward.} transitions from some given initial state $ \Psi $ with energy $ E $ to a lower state $ \Psi' $ with energy $ E' $. Assuming the system of the wave functions is orthonormalized, the total radiation power obtained using the methods of QED (the standard Feynman rules) can be written in the first order of $ e^{2} $ as follows (see, e.g., \cite{book_qed}, and also \cite{book_rel_el}, \cite{book_hep_rad}; we follow the approach of \cite{book_rel_el}, \cite{book_hep_rad} while describing the theory of the synchrotron radiation in this section):
\begin{equation}
    \label{rad_W_def}
    W = \frac{e^{2}}{2\pi} \int d^{3}k \ \delta( E - E' - k ) \, S, \qquad
    S = \left| \la \bs \alpha \ra {\bf f} \right|^{2}.
\end{equation}
Here $ {\bf k} $ is the wave vector of the photon emitted, so that the energy of the photon is $ \omega = k \equiv |{\bf k}| $; $ {\bf f} $ is the vector characterizing the polarization properties of the photon (it is always orthogonal to $ {\bf k} $; the radiation is treated in the temporal gauge); the vector quantity $ \la \bs \alpha \ra $ is related to the transition amplitude:
\begin{equation}
    \label{rad_alpha_def}
    \la \bs \alpha \ra =
        \int d^{3}x \, \Psi'^{\dag} \left( \bs \alpha e^{-i {\bf kx}} \right) \Psi.
\end{equation}

Let $ (\theta, \varphi) $ be the angles characterizing the direction of the radiation of a given
polarization in a spherical coordinate system with the $ z $-axis parallel to the magnetic
field orientation, so that
\begin{equation}
    \label{rad_k}
    {\bf k} =
        k \{ \sin \theta \cos \varphi, \sin \theta \sin \varphi, \cos \theta \}.
\end{equation}
Evaluating the integral in (\ref{rad_alpha_def}), due to the general form of the wave functions (\ref{sol_Psi_phi}), (\ref{sol_phi}), one finds (see \cite{book_hep_rad} for the details of these calculations):
\begin{equation}
    \la \bs \alpha \ra \ =
        \la \bar{\bs \alpha} \ra \, I_{s,s'}(x) \, \delta( p' - p + k \cos \theta ),
\end{equation}
where
\begin{equation}
    \label{rad_alpha}
    \begin{aligned}
        -i \la \bar{\alpha}_{1} \ra \ & = &&
            ( c_{1}'^{*} c_{4}^{} + c_{3}'^{*} c_{2}^{} ) I_{n,n'-1}(x)
            && - && ( c_{1}^{} c_{4}'^{*} + c_{3}^{} c_{2}'^{*} ) I_{n-1,n'}(x), \\
        \la \bar{\alpha}_{2} \ra \ & = &&
            ( c_{1}'^{*} c_{4}^{} + c_{3}'^{*} c_{2}^{} ) I_{n,n'-1}(x)
            && + && ( c_{1}^{} c_{4}'^{*} + c_{3}^{} c_{2}'^{*} ) I_{n-1,n'}(x), \\
        \la \bar{\alpha}_{3} \ra \ & = &&
            ( c_{1}'^{*} c_{3}^{} + c_{3}'^{*} c_{1}^{} ) I_{n-1,n'-1}(x)
            && - && ( c_{2}^{} c_{4}'^{*} + c_{4}^{} c_{2}'^{*} ) I_{n,n'}(x), \\
    \end{aligned}
\end{equation}
and this is obtained, of course, using the standard representation of the $ \gamma $-matrices we have chosen. The argument of all Laguerre functions in (\ref{rad_alpha}) is defined as follows:
\begin{equation}
    x = \frac{1}{2eH} \, k^{2} \sin^{2} \theta.
\end{equation}
The quantities with and without a dash in (\ref{rad_alpha}) are related to the final and initial state of the electron respectively, and we also use this notation in what follows.

Making the summation over the quantum numbers $ n', s', p' $ characterizing the final state and evaluating the integral over $ k $ in (\ref{rad_W_def}), one obtains the expression for the radiation power (related to one unit of length of the $ z $-axis):
\begin{equation}
    \label{rad_W_sum}
    W =
        \sum_{n'} \frac{e^{2}}{2\pi}
        \int \frac
        { k^{2} \sin \theta \, d\theta \, d\varphi }
        { \left|
            1 + \left( \frac{\partial E'}{\partial k} \right)_{n'}
        \right | } \,
        \bar{S}, \qquad
        \bar{S} = \left| \la \bar{\bs \alpha} \ra {\bf f} \right|^{2},
\end{equation}
where $ k $ and $ p' $ obey the conservation laws of the energy and $ z $-component of the momentum:
\begin{equation}
    \label{rad_conserv_cond}
    E' = E - k, \quad p' = p - k \cos \theta.
\end{equation}
We use the symbol $ \left( \frac{\partial E'}{\partial k} \right)_{n'} $ to denote a derivative of the energy $ E' $ considered as a function of $ p' $, where $ p' $ is defined through (\ref{rad_conserv_cond}), with respect to $ k $, with $ n' $ being fixed. It is important that since there exists the relation (see \cite{book_hep_rad}):
\begin{equation}
    \sum_{s' = 0}^{\infty} I_{s,s'}(x) I_{m,s'}(x) = \delta_{sm} \qquad
    \text{(for all allowed $ x $)},
\end{equation}
the summation has taken the numbers $ s $, $ s' $ out of consideration; this is closely connected with the degeneracy and invariance existing in our problem (see Section \ref{sol}). Thus, the initial quantum number $ s $ may be arbitrary.

Note that we are still considering the electron initial and final states with the definite spin quantum numbers $ \zeta $ and $ \zeta' $ respectively, i.e., we do not make any averaging or summation over them.




    \subsection{Radiation of an ultra-relativistic electron}

    Let us now
consider the most interesting case of a high-energy particle ($ m / E \equiv \lambda \ll 1 $) in a comparatively weak magnetic field ($ H \ll H_{c} = m^{2} / e \simeq 4.41 \cdot 10^{13} \, \text{gauss} $) with the initial longitudinal momentum $ p = 0 $, which corresponds to the electron 
states with $ n \gg 1 $.
    Indeed, examining this case, one approximately finds from (\ref{sol_E}):
    \begin{equation}
        \label{rad_n_est}
        n \simeq \frac{1}{2 \lambda^{2}} \frac{H_{c}}{H} \gg 1.
    \end{equation}
When calculating radiation effects we shall 
restrict ourselves to the zero approximation in $ \tilde{\mu}H / E $. In fact, it is not difficult to prove that only three small parameters are of importance in our problem: $ \lambda $, $ \delta $, $ \tilde{\mu}H / E $. However, it is easily seen then that under typical laboratory conditions ($ E \sim 1 \, \text{GeV}$, $ H \sim 10^{4} \, \text{gauss} $) the estimate $ | \delta | \gg \tilde{\mu}H / E $
is valid if only $ b \gg 10^{-20} \, \text{eV} $, which justifies our approximation $ \tilde{\mu}H / E \rightarrow 0 $ for this range of $ b $
    (see also the Discussion).

It is obvious that the chosen approximation $ \tilde{\mu} \rightarrow 0 $ reduces the problem to
the case of the Dirac hamiltonian
\begin{equation}
    \label{rad_H}
    H^{0}_{\rm D} =
        \bs \alpha {\bf P} + \gamma^{0} m
\end{equation}
which follows from (\ref{model_H}), when $ \mu \rightarrow 0 $, $ b \rightarrow 0 $. Operator
(\ref{rad_H}) has the spectrum
\begin{equation}
    \label{rad_E}
    E = \sqrt{
        m^{2} + 2eHn + p^{2}
    },
\end{equation}
and admits of 
an ambiguity in the choice of a spin operator commuting with it (see, e.g., \cite{book_rel_el}, \cite{book_hep_rad}). In our case, however, this ambiguity is removed and the operator (\ref{sol_Pi_def}) should be used, which corresponds to a ``mixed'' (transversal-longitudinal) polarization of the particle. Thus, the problem is formally reduced
to the radiation of an electron without an anomalous magnetic moment or Lorentz symmetry violation, polarized in a definite way. This (namely the dispersion law (\ref{rad_E}) and the general form of the wave functions (\ref{sol_Psi_phi}), (\ref{sol_phi})) makes the well-known
theory of synchrotron radiation considered in \cite{book_rel_el}, \cite{book_hep_rad} applicable to our case.


In particular, if relation (\ref{rad_E}) holds, then system (\ref{rad_conserv_cond}) can be solved unambiguously with respect to $ k $, $ p' $, and this yields (note that we are considering the case $ p = 0 $):
\begin{equation}
    \label{rad_k_from_conserv_cond}
    k = \frac{E}{\sin^{2} \theta} \left(
        1 - \sqrt{ 1 - \beta^{2} \left( 1 - \frac{n'}{n} \right) \sin^{2} \theta }
    \right), \qquad
    \beta^{2} = 1 - \lambda^{2}.
\end{equation}
Since we are considering the states with $ n \gg 1 $, it is a good approximation to change the sum in (\ref{rad_W_sum}) into an integral treating $ n' $ as a continuous variable. With the help of (\ref{rad_k_from_conserv_cond}), it is possible to change the variable of integration from $ n' $ to $ k $ explicitly:
\begin{equation}
    \frac{ dn' }{\left| 1 + \left( \frac{\partial E'}{\partial k} \right)_{n'} \right|} =
        - \frac{E'}{eH} \, dk.
\end{equation}
In that way, one obtains the spectral-angular distribution $ w^{(k)} $ of the radiation (we imply, of course, that $ 0 < k < E $):
\begin{equation}
    \label{rad_W_w_with_k}
    W = \int dk \, \sin \theta \, d\theta \, d\varphi \, w^{(k)}, \qquad
    w^{(k)} = \frac{e^2}{2\pi} \frac{E' k^{2}}{eH} \, \bar{S}.
\end{equation}

In the case $ n, n' \gg 1 $ we are interested in\footnote{It can be seen that transitions to the states with $ n' \sim 1 $ are actually suppressed when $ n \gg 1 $, so that our consideration is consistent. The issues concerning the approximation we make are discussed in \cite{book_rel_el} in more detail.}, there exist the asymptotic expressions for the Laguerre functions present in (\ref{rad_alpha}):
\begin{equation}
    \label{rad_I_K}
    \begin{pmatrix}
        I_{n,n'}(x)     \\
        I_{n,n'-1}(x)   \\
        I_{n-1,n'}(x)   \\
        I_{n-1,n'-1}(x) \\
    \end{pmatrix} \simeq
    \eta \left\{
        \tilde{\lambda} K_{1/3}(z) +
        \begin{pmatrix}
            0            \\
            -(1 + \xi y) \\
            1            \\
            - \xi y      \\
        \end{pmatrix}
        \tilde{\lambda}^{2} K_{2/3}(z) +
        {\cal O}(\lambda^3)
    \right\},
\end{equation}
where $ K_{\nu}(z) $ are the modified Bessel functions of the second kind,
\begin{equation}
    \label{rad_I_K_vars}
    \begin{aligned}
        \eta = \sqrt{\frac{1 + \xi y}{3\pi^{2}}}, \quad & \quad
        \xi = \frac{3}{2} \frac{H}{H_{c}} \frac{1}{\lambda}, \\
        z =
            \frac{y}{2} \left( \frac{\tilde{\lambda}}{\lambda} \right)^{3}
            \left( 1 + {\cal O}(\lambda^2) \right), \quad & \quad
        \tilde{\lambda}^{2} = 1 - \beta^{2} \sin^{2} \theta, \\
    \end{aligned}
\end{equation}
The dimensionless spectral variable $ y $ is related to the photon energy $ k $ by the formula:
\begin{equation}
    \label{rad_y_def}
    \frac{k}{E} =
        \frac{\xi y}{1 + \xi y}, \qquad
    0 < y < +\infty.
\end{equation}
The coefficients in front of the functions $ K_{\nu}(z) $ in (\ref{rad_I_K}) and the quantity $ z $ in (\ref{rad_I_K_vars}) are written in the second and leading order of $ \lambda $ respectively taking into account that, as it can be easily seen, the quantities $ \tilde{\lambda} $ and $ \cos \theta $ also have the same order of smallness as $ \lambda $. The latter is valid in the range of $ \theta $ where $ K_{\nu}(z) $ are essentially different from zero; from the physical point of view this corresponds to the fact that the radiation of a high-energy particle is concentrated close to the plane of its orbit, and the typical angle of deviation or the spread angle is $ \Delta \theta \simeq \lambda $.

Now using the new variable $ y $ instead of $ k $, one finds:
\begin{equation}
    \label{rad_W_w_with_y}
    W = W_{cl} \int dy \, \sin \theta \, d\theta \, d\varphi \, w^{(y)}, \qquad
    w^{(y)} =
        \frac{27}{128 \pi^{3}} \, \frac{y^{2}}{\lambda^{5} (1 + \xi y)^{4}} \, \Phi,
\end{equation}
where $ W_{cl} = \frac{8}{27} \, e^{2} m^{2} \xi^{2} $ is the total power of the synchrotron radiation of a high-energy particle in the classical limit and $ w^{(y)} $ is the spectral-angular distribution (normalized and related to $ y $). We shall omit the superscript and denote it as $ w $ in what follows. We also express it through the quantity $ \Phi = 4 \bar{S} / \eta^{2} $, since $ \Phi $ represents the angular distribution in a rather convenient way (see the results below).

The dimensionless quantity $ \xi $ introduced in (\ref{rad_I_K_vars}) is known as the parameter characterizing the quantum corrections to the radiation. Actually, recovering the dimensional constants $ \hbar $ and $ c $, we find (note that we are still using the natural scale for $ e $ so that the fine structure constant is $ \alpha = e^{2} / 4 \pi \hbar c $):
\begin{equation}
    \xi = \frac{3 \, e H E}{2 \, m^{3}} \rarr
        \frac{3 \, e \hbar \, H E}{2 \, m^{3} c^{5}} \sim \hbar.
\end{equation}
It is easily seen that $ \hbar $ emerges exactly through $ \xi $ in our problem. At the same time, $ \hbar $ is cancelled out in the leading order (classical) expressions, e.g.,
\begin{equation}
    W_{cl} = \frac{8}{27} \, e^{2} m^{2} \xi^{2} \rarr
        \frac{8}{27} \frac{e^{2} m^{2} c^{3}}{\hbar^{2}} \, \xi^{2} =
            \frac{2 e^{2}}{3 c}
            \left( \frac{eH}{mc} \right)^{2}
            \left( \frac{E}{mc^{2}} \right)^{2}.
\end{equation}
In the case under investigation, although we require $ \lambda \ll 1 $, $ H \ll H_{c} $, the quantity $ \xi $ may take {\it arbitrary} values, and we do not make any expansion in its powers. In this respect, our results are exact, including all the quantum corrections arising. At the same time, pure classical methods widely adopted in the literature are only applicable when $ \xi \ll 1 $.


\subsection{Spectral-angular distribution}

Exploiting (\ref{rad_I_K}), one can obtain the corresponding asymptotic expressions for $ \bar{S} $, $ \Phi $ and thus for $ w $. Considering the $ \sigma $- and $ \pi $-components of the linear polarization of radiation\footnote{In the case of the $ \sigma $-polarization, the electric field vector $ {\bf E} $ is in the $ (xy) $-plane, while for the $ \pi $-polarization, the magnetic field vector {\bf H} lies in this plane.}, one can choose $ {\bf f} $ as follows (see, e.g., \cite{book_hep_rad}):
\begin{equation}
    \label{rad_f_sigma_pi}
    \begin{aligned}
        {\bf f}_{\sigma} & = \{ 1, 0, 0 \}, \\
        {\bf f}_{\pi}    & = \{ 0, \cos \theta, -\sin \theta \}, \\
    \end{aligned}
\end{equation}
this corresponds to $ \varphi = \frac{\pi}{2} $ in (\ref{rad_k}). The specific value of the angle $ \varphi $ is inessential due to the axial symmetry present in our problem. Thus, one obtains:
\begin{equation}
    \label{rad_S_sigma_pi}
    \begin{aligned}
        \bar{S}_{\sigma} & = \left| \la \bar{\alpha}_{1} \ra \right|^{2}, \\
        \bar{S}_{\pi}    & =
            \left| \la \bar{\alpha}_{2} \ra \cos \theta
            - \la \bar{\alpha}_{3} \ra \sin \theta \right|^{2}. \\
    \end{aligned}
\end{equation}
According to (\ref{rad_conserv_cond}) and (\ref{rad_y_def}) (taking into account that $ p = 0 $), one has:
\begin{equation}
    \label{rad_lambda_pbar_fin}
    \frac{m}{E'} \equiv \lambda' = \lambda (1 + \xi y), \quad
    \frac{p'}{E'} \equiv \bar{p}' = -\xi y \cos \theta,
\end{equation}
this implies that $ \lambda' $, $ \bar{p}' $ also have the order of smallness of $ \lambda $. Note that, by means of (\ref{sol_Pi_values}), (\ref{rad_E}) the quantity $ \Pi $ can be expressed in terms of $ E $ and $ \tilde{p} $ as follows:
\begin{equation}
    \label{rad_Pi}
    \Pi =
        \zeta \sqrt{E^{2} - \tilde{p}^{2}}.
\end{equation}
Deriving the latter formula, we have neglected the case $ n = 0 $ in (\ref{sol_Pi_values}) since it is inessential in our approach. Now it is possible to write down the quantities defined in (\ref{sol_ABPQ}), related both to the initial and the final states, expressing them in terms of $ \lambda $, $ \cos \theta $, $ \lambda' $, $ \bar{p}' $ (and also $ \zeta $, $ \zeta' $). So we use (\ref{rad_alpha}), then (\ref{sol_coeffs}), (\ref{sol_alpha_beta}), (\ref{sol_ABPQ}) (taking into account (\ref{rad_Pi})) to evaluate (\ref{rad_S_sigma_pi}); after that we make expansions in powers of $ \lambda $, $ \cos \theta $, $ \lambda' $, $ \bar{p}' $, keeping the terms up to the order of $ \lambda^{2} $. Finally, we use the asymptotic formulae (\ref{rad_I_K}), preserving only the leading order in $ \lambda $. It appears that the accuracy of the expansions we made is necessary and sufficient for carrying out the calculations consistently.

Without going into details of the calculations described, we present here the final result for $ \Phi_{i} $, $ i = \sigma, \pi $ (having expressed $ \lambda' $, $ \bar{p}' $ through  $ \lambda $, $ \cos \theta $ according to (\ref{rad_lambda_pbar_fin})). Separating the transitions with and without the change of the spin quantum number, one obtains:
\begin{equation}
    \Phi_{i} =
        \frac{1 + \zeta \zeta'}{2} \Phi^{+}_{i} + \frac{1 - \zeta \zeta'}{2} \Phi^{-}_{i},
\end{equation}
where
\begin{equation}
    \label{rad_Phi_sigma}
    \setlength\arraycolsep{2pt}
    \begin{array}{rcl}
        \Phi_{\sigma}^{+} & = &
            \tilde{\lambda}^{2}
            \Bigl(
                (2 + \xi y) \tilde{\lambda} K_{2/3}(z)
                - \zeta (\xi y) (\lambda \cos \delta - \cos \theta \sin \delta) K_{1/3}(z)
            \Bigl)^{2}, \\
        \Phi_{\sigma}^{-} & = &
            \tilde{\lambda}^{2}
            \Bigl(
                (\xi y) (\cos \theta \cos \delta + \lambda \sin \delta) K_{1/3}(z)
            \Bigl)^{2}, \\
    \end{array}
\end{equation}
and
\begin{equation}
    \label{rad_Phi_pi}
    \setlength\arraycolsep{2pt}
    \begin{array}{rcl}
        \Phi_{\pi}^{+} & = &
            \tilde{\lambda}^{2}
            \Bigl(
                (2 + \xi y) \cos \theta \, K_{1/3}(z)
                + \zeta (\xi y) \sin \delta \, \tilde{\lambda} K_{2/3}(z)
            \Bigl)^{2}, \\
        \Phi_{\pi}^{-} & = &
            \tilde{\lambda}^{2}
            \Bigl(
                (\xi y) \bigl(
                    \cos \delta \, \tilde{\lambda} K_{2/3}(z) + \zeta \lambda K_{1/3}(z)
                \bigl)
            \Bigl)^{2}. \\
    \end{array}
\end{equation}
When $ \delta = 0, \, \pi / 2 $, formulae
(\ref{rad_Phi_sigma})--(\ref{rad_Phi_pi}) turn obviously into the well-known ones from the synchrotron radiation theory of a transversally and longitudinally polarized electron (see, e.g., \cite{book_hep_rad}).

    %

\begin{figure}[t]
    \noindent
    \centering
    \footnotesize
    $
    \begin{array}{rr}
        \tilde{\Phi}^{+}_{\sigma}(\theta), \ \zeta = -1; \ a \simeq -1.2 \cdot 10^{-8} &
        \tilde{\Phi}^{-}_{\sigma}(\theta), \ \zeta = -1; \ a \simeq 0.11 \\
        \includegraphics[width=0.475\textwidth]{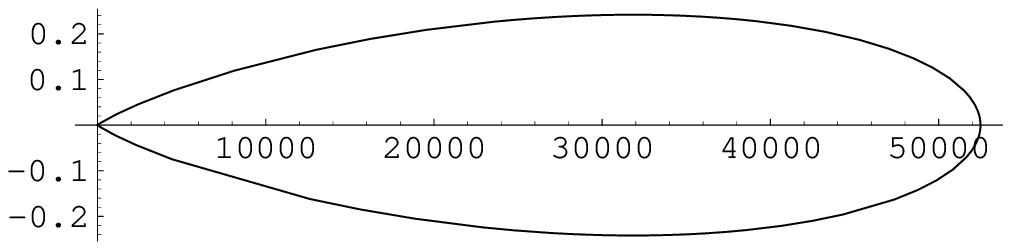} &
        \includegraphics[width=0.475\textwidth]{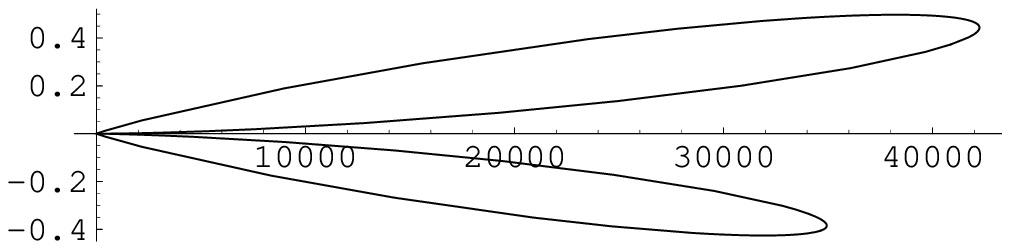} \\
        & \\
        \tilde{\Phi}^{+}_{\pi}(\theta), \ \zeta = -1; \ a \simeq -5.4 \cdot 10^{-5} &
        \tilde{\Phi}^{-}_{\pi}(\theta), \ \zeta = -1; \ a \simeq 0 \\
        \includegraphics[width=0.475\textwidth]{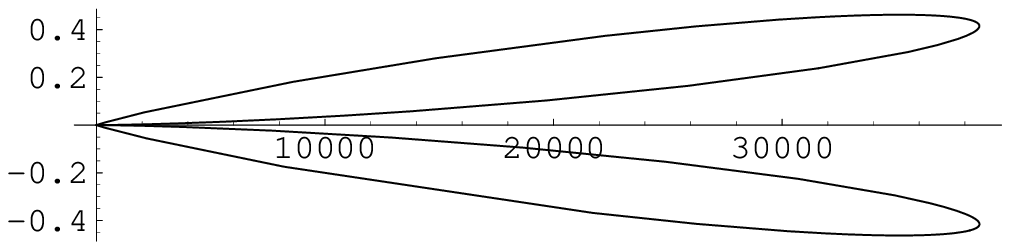} &
        \includegraphics[width=0.475\textwidth]{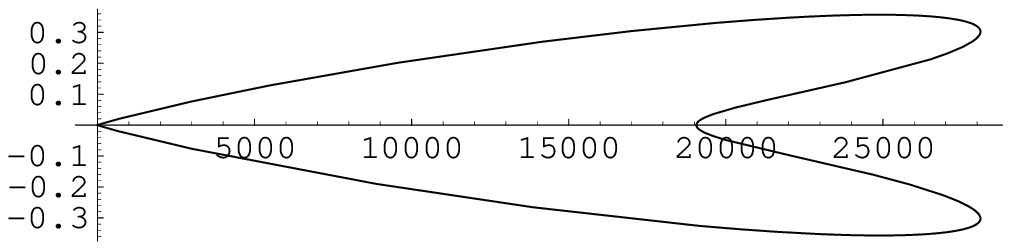} \\
        & \\
        \tilde{\Phi}^{+}_{\sigma}(\theta), \ \zeta = +1; \ a \simeq 1.2 \cdot 10^{-8} &
        \tilde{\Phi}^{-}_{\sigma}(\theta), \ \zeta = +1; \ a \simeq 0.11 \\
        \includegraphics[width=0.475\textwidth]{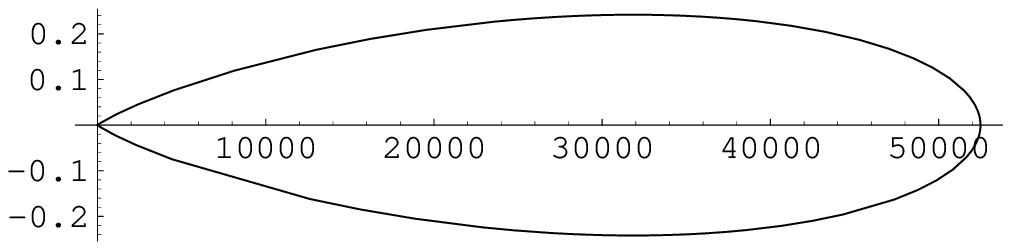} &
        \includegraphics[width=0.475\textwidth]{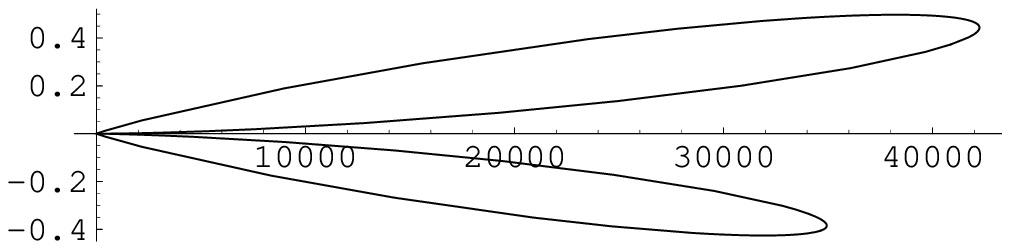} \\
        & \\
        \tilde{\Phi}^{+}_{\pi}(\theta), \ \zeta = +1; \ a \simeq 5.4 \cdot 10^{-5} &
        \tilde{\Phi}^{-}_{\pi}(\theta), \ \zeta = +1; \ a \simeq 0 \\
        \includegraphics[width=0.475\textwidth]{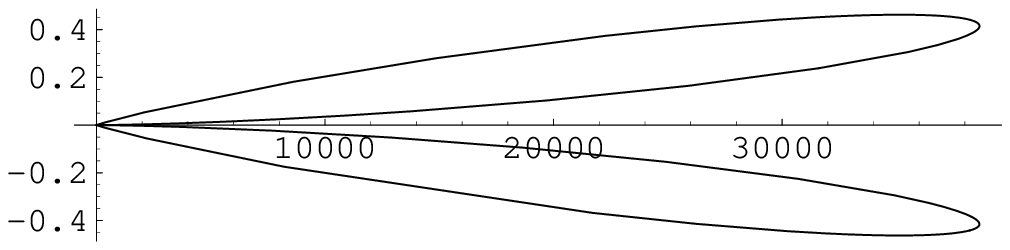} &
        \includegraphics[width=0.475\textwidth]{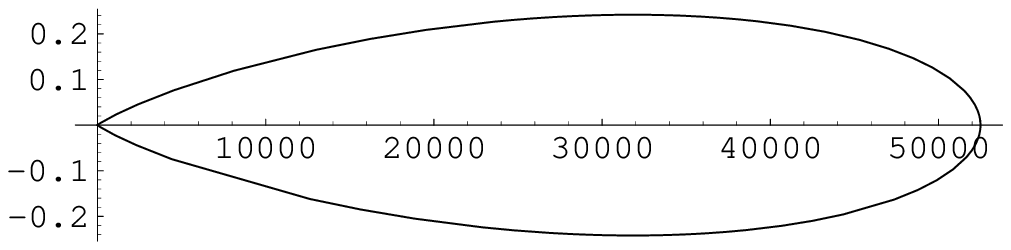} \\
    \end{array}
    $
    \caption{
        The normalized angular distributions $ \tilde{\Phi}^{\pm}_{i}(\theta) $
        in the polar coordinate system $ (\tilde{\Phi},\theta) $
        plotted for $ k = 1 \, \text{MeV} $, $ \zeta = \pm 1 $,
        and $ H = 10^{4} \, \text{gauss} $, $ E = 1 \, \text{GeV} $, $ \delta = 10^{-3} $.
    }
    \label{fig:Phi_phys}
\end{figure}

\begin{figure}[t]
    \noindent
    \centering
    \footnotesize
    $
    \begin{array}{rr}
        \tilde{\Phi}^{+}_{\sigma}(\theta), \ \zeta = -1; \ a \simeq -0.011 &
        \tilde{\Phi}^{-}_{\sigma}(\theta), \ \zeta = -1; \ a \simeq 0.19 \\
        \includegraphics[width=0.475\textwidth]{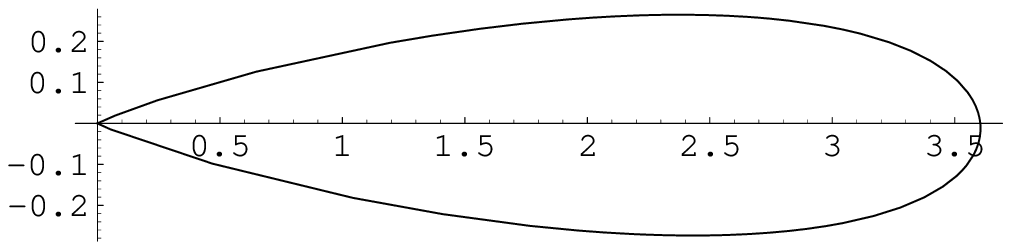} &
        \includegraphics[width=0.475\textwidth]{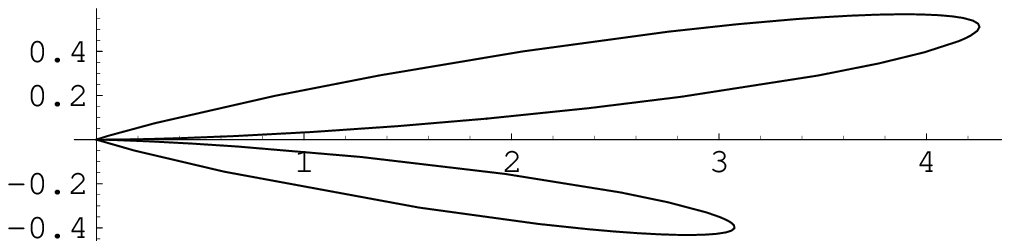} \\
        & \\
        \tilde{\Phi}^{+}_{\pi}(\theta), \ \zeta = -1; \ a \simeq -0.052 &
        \tilde{\Phi}^{-}_{\pi}(\theta), \ \zeta = -1; \ a \simeq 0 \\
        \includegraphics[width=0.475\textwidth]{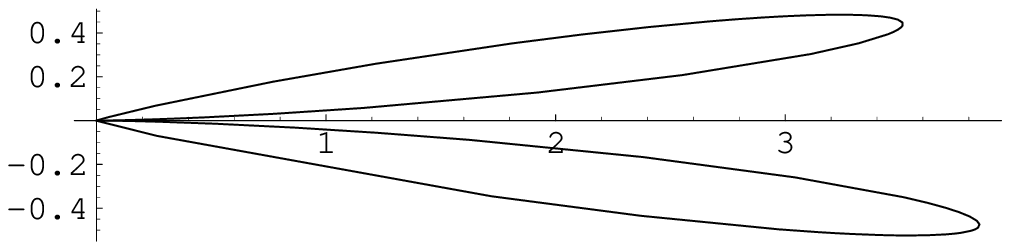} &
        \includegraphics[width=0.475\textwidth]{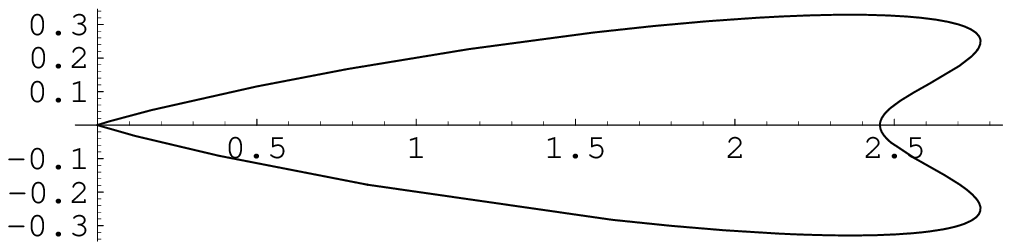} \\
        & \\
        \tilde{\Phi}^{+}_{\sigma}(\theta), \ \zeta = +1; \ a \simeq 0.013 &
        \tilde{\Phi}^{-}_{\sigma}(\theta), \ \zeta = +1; \ a \simeq 0.19 \\
        \includegraphics[width=0.475\textwidth]{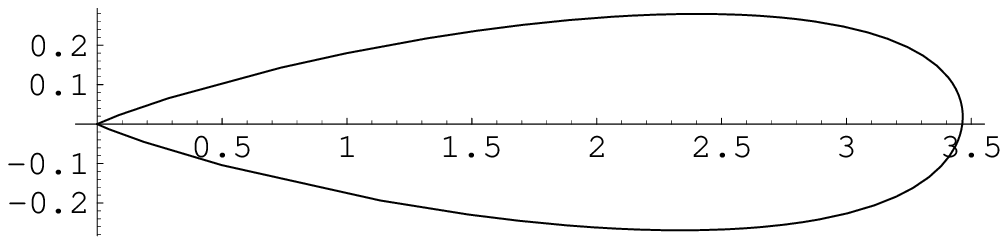} &
        \includegraphics[width=0.475\textwidth]{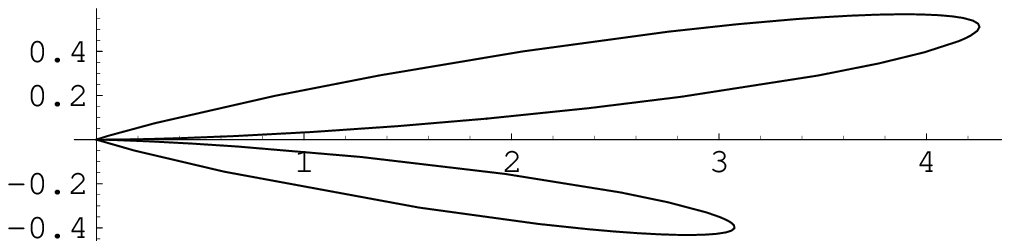} \\
        & \\
        \tilde{\Phi}^{+}_{\pi}(\theta), \ \zeta = +1; \ a \simeq 0.052 &
        \tilde{\Phi}^{-}_{\pi}(\theta), \ \zeta = +1; \ a \simeq 0 \\
        \includegraphics[width=0.475\textwidth]{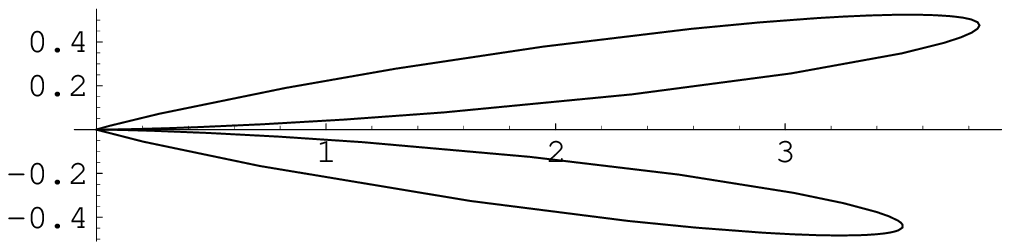} &
        \includegraphics[width=0.475\textwidth]{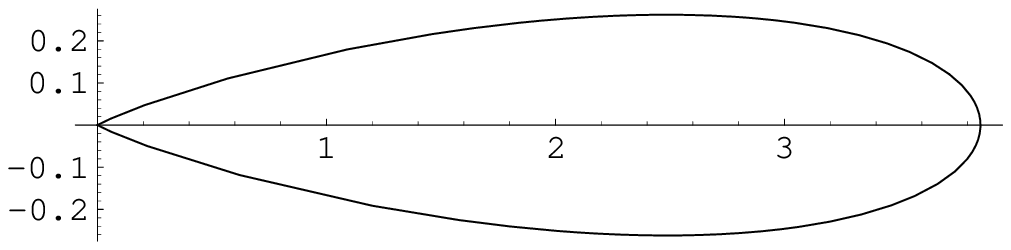} \\
    \end{array}
    $
    \caption{
        The normalized angular distributions $ \tilde{\Phi}^{\pm}_{i}(\theta) $
        in the polar coordinate system $ (\tilde{\Phi},\theta) $
        plotted for $ k = 0.25 \, E $, $ \zeta = \pm 1 $,
        and $ H = 0.1 \, H_{c} $, $ E = 10 \, m $, $ \delta = 0.1 $.
    }
    \label{fig:Phi_good}
\end{figure}

From (\ref{rad_Phi_sigma})--(\ref{rad_Phi_pi}) one can see that the effect caused by the proposed existence of Lorentz invariance violation in our theory reveals itself in asymmetry of the angular
distributions $ \Phi^{\pm}_{i}(\theta) $ relative to the $ \theta = \pi / 2 $ plane (i.e. to the plane of the particle orbit). Asymmetry of this kind is inherent in the radiation of the longitudinally polarized electron and is totally absent in the case of the transversal polarization; one should remember that considering the influence of the anomalous magnetic moment (without Lorentz invariance violation) only the transversal polarization of an electron moving in a magnetic field is conserved. This fact is obvious from (\ref{sol_Pi_def}) with $ \delta = 0 $, of course, in case when the states of an electron under consideration are the eigenvectors of its hamiltonian and the time dependence of its wave functions is described by the factor $ e^{-iEt} $ (for further details see \cite{book_rel_el}). Thus, if Lorentz invariance violation is present (in the form we have chosen in this work) the electron spin integral of motion receives an additional longitudinal
part and takes the form (\ref{sol_Pi_def}), and according to this,
the electron electromagnetic radiation is changed.

Experimental detection of asymmetry of the angular distribution of the synchrotron radiation of
electrons polarized according to their own (conserved) spin operator could provide a possibility
to estimate the quantity $ \delta $ and hence $ b $ as its function. In order to characterize this asymmetry one can use, e.g., the quantity
\begin{equation}
    a = \frac{w_{up} - w_{down}}{w_{up} + w_{down}},
\end{equation}
where
\begin{equation}
    w_{up} =
        \int_{0}^{\frac{\pi}{2}} \sin \theta \, d\theta \, \Phi, \quad
    w_{down} =
        \int_{\frac{\pi}{2}}^{\pi} \sin \theta \, d\theta \, \Phi.
\end{equation}
It is clear that in the linear approximation we have
$ a \sim \delta $, where the proportionality factor can be numerically calculated for any given photon energy. The typical curves for $ \Phi^{\pm}_{i}(\theta) $ (namely, for the normalized functions $ \tilde{\Phi}^{\pm}_{i}(\theta) = \Phi^{\pm}_{i}(\theta) / N $, where $ N = \int_{0}^{\pi} \sin \theta \, d\theta \, \Phi^{\pm}_{i}(\theta)$) are depicted in Figs. \ref{fig:Phi_phys} and \ref{fig:Phi_good}. The corresponding asymmetry coefficient $ a $ is shown in each diagram. In Fig. \ref{fig:Phi_good} we plotted the curves for the high values of $ H $, $ \delta $ and the low value of $ E $ in order to make the asymmetry in $ \Phi^{+}_{i}(\theta) $ more visible.


    \section{Discussion and conclusions}
\label{outro}

The change in the angular distribution and the specific asymmetry of the synchrotron radiation of an electron caused by the assumed existence of Lorentz invariance violation have already been discussed in literature,
see, e.g., \cite{Montemayor_Urrutia}. However, calculations in
\cite{Montemayor_Urrutia}, as well as in
\cite{Ellis_Mavromatos_Sakharov}, were based on semi-classical methods outside the framework of the Standard Model Extension. Moreover, specific modified Lorentz non-invariant dispersion laws for photons and electrons were adopted there. In
this paper, in contrast to \cite{Montemayor_Urrutia}, \cite{Ellis_Mavromatos_Sakharov}, we used the SME technique and employed the standard methods of QED. Our results are based on exact solutions of the Dirac equation for an electron with a vacuum magnetic moment in a constant magnetic field. They are due to the peculiar non-perturbative interaction of the electron vacuum magnetic moment $ \mu $ with the condensate $ b^{\alpha} $ violating Lorentz invariance.

    In our work, we considered the radiation phenomena at the entirely quantum level and obtained the asymptotic expressions for the spectral-angular distribution for the case of a high-energy particle moving in a relatively weak magnetic field; these expressions incorporate all the quantum corrections in our problem. The reason is that specific effects caused by the Lorentz invariance violation present in the theory are closely connected with spin effects, and they should be handled using the quantum approach; moreover, classical and semi-classical methods widely adopted in the literature (see, e.g., \cite{Altschul}, and also \cite{Montemayor_Urrutia}, \cite{Ellis_Mavromatos_Sakharov}) may not be actually applicable.

    The results we have obtained provide us with the possibility of making some estimates of the parameter $ b $ governing the effects of Lorentz invariance violation in our problem. First of all, it is believed (see, e.g., \cite{book_hep_rad} and references therein) that it is the transversal electron polarization that is primarily observed in synchrotron-radiation experiments. This implies that the ``mixing angle'' $ \delta $ defined in (\ref{sol_vartheta_def}) should be sufficiently small (since the type of polarization depends on it, see (\ref{sol_Pi_def})) under the laboratory conditions ($ E \sim 1 \, \text{GeV}$, $ H \sim 10^{4} \, \text{gauss}$), and this yields the estimate:
    \begin{equation}
        \label{outro_b_est}
        | b | \ll \mu H \sim 10^{-6} \, \text{eV}.
    \end{equation}
    In the first approximation, we used here well-known Schwinger's result for the value of the vacuum magnetic moment: $ \mu \simeq \mu_{0} \, \frac{e^{2}}{2\pi} $, where $ \mu_{0} = \frac{e}{2m} $ (see also the discussion below). Result (\ref{outro_b_est}) is more stringent than the typical estimates available in the literature, e.g., in \cite{Andrianov_Giacconi_Soldati}.

    At the same time, one may argue that in case of reliable observation of radiation of the vacuum magnetic moment (the theory of which was developed in \cite{Ternov_Bagrov_Zhukovsky}) in the absence of any effects caused by $ \delta \neq 0 $, the estimate becomes appreciably stronger. In fact, as it has been shown, the prevailing effect of $ \delta \neq 0 $ is the asymmetry of the synchrotron radiation, while the radiation of the vacuum magnetic moment (governed by the small parameter $ \mu H / E $) is symmetric with respect to the plane of the particle orbit (of course, it is assumed here that the particle is transversally polarized and no Lorentz invariance violation is then present). So if no asymmetry is detected but the radiation of the vacuum magnetic moment is actually observed (and it is not distorted in any way), it is reasonable to consider that the small parameters of our problem are related as follows: $ | \delta | \ll \tilde{\mu}H / E $. By virtue of the definitions of $ \delta $ and $ \tilde{\mu} $ (\ref{sol_vartheta_def}), (\ref{sol_mueff_def}), this condition also guarantees $ \tilde{\mu}H / E \simeq \mu H / E $ with very good precision, and it is satisfied when
    \begin{equation}
        \label{outro_b_est_cool}
        | b | \ll \frac{(\mu H)^{2}}{E} \sim 10^{-20} \, \text{eV} .
    \end{equation}
    However, the latter estimate may need more justification since we did not provide explicit calculations, and there are also some issues concerning the possibility of reliable experimental observation of radiation of the vacuum magnetic moment and its interpretation. The fact is, when $ b \gg 10^{-20} \, \text{eV} $ we may neglect the small parameter $ \tilde{\mu}H / E $ while considering the radiation phenomena.

In our calculations we regarded the quantity $ \mu $ as a constant, which depends neither on the particle state nor on the magnetic field strength. However, consistent consideration of the vacuum
magnetic moment phenomenon leads to the conclusion that such a dependence does exist, and, in particular, $ \mu $ decreases with growing particle energy \cite{book_rel_el}, \cite{book_hep_rad}. 
    At the same time, the existence of Lorentz invariance violation, of course, also modifies the vacuum magnetic moment behavior \cite{sme_vmm_theory}. When considering this issue, one has to work with not only the fermion but also the photon sector of the model due to the nature of the vacuum magnetic moment phenomenon \cite{sme_vmm_from_phot}. Nonetheless, in our case, at experimentally feasible energies, the vacuum magnetic moment should have the general form:
    \begin{equation}
        \label{outro_mu_liv}
        \mu = \mu^{\rm std} + {\text{\{corrections\}}},
    \end{equation}
    where $ \mu^{\rm std} $ denotes the well-known standard result for the function $ \mu(\frac{E}{m},\frac{H}{H_{c}}) $ \cite{book_rel_el}, \cite{book_hep_rad} (in the zero approximation, in a weak field, it is the constant value found by Schwinger), and the corrections arise through Lorentz non-invariant interactions of the particle. Thus, one has from (\ref{sol_vartheta_def}) and (\ref{outro_mu_liv}):
    \begin{equation}
        \delta = \arctan \left( \frac{b}{\mu^{\rm std} H + \ldots} \right),
    \end{equation}
    and since we are actually interested in calculating $ \delta $ in the leading order of the parameters characterizing Lorentz invariance violation (as we believe that they are {\it small} numbers), we may neglect these additional corrections, preserving only $ b $ in the numerator:
     \begin{equation}
        \delta \simeq \frac{b}{\mu^{\rm std} H}.
    \end{equation}
    The latter expression is finally used when deriving all the estimates.

Thus, according to the comments given above
and to definition (\ref{sol_vartheta_def}) of the angle $ \delta $, one may conclude that it should increase with growing electron energy (while the second small parameter $ \tilde{\mu}H / E $ of our problem, on the contrary, decreases at least like $ 1 / E $). This means that the effect of the asymmetry of the angular distribution of the synchrotron radiation described in this work should become more noticeable with growing electron energy $ E $.

It should be noted, however, that since formulae (\ref{rad_Phi_sigma})--(\ref{rad_Phi_pi}) of this
work were obtained under the assumption $ \delta = \const $, one can use them for relatively soft
photons only, when final electron states have approximately the same value of $ \delta $ as the
initial one.

The type of the Lorentz symmetry breaking studied in this paper on the basis of the Standard Model Extension and the particular choice of $ b^{\mu} = \{ b, {\bf 0} \} $ does not, of course, exhaust all the possibilities, that can actually find their realization in nature. Nonetheless, the effect of mixing of transversal and longitudinal polarizations of an electron, with an interpretation suggested in this work, may provide one of the tests of the possible violation of Lorentz invariance.

\section*{Acknowledgements}

The authors are grateful to A.~V.~Borisov, D.~Ebert, and A.~E.~Lobanov for useful discussions.




\begin{thebibliography}{99}

    \bibitem{sme_theory}
    S. M. Carroll, G. B. Field, and R. Jackiw, Phys. Rev. D {\bf 41}, 1231 (1990); D. Colladay and V. A. Kosteleck\'{y}, Phys. Rev. D {\bf 55}, 6760 (1997), hep-ph/9703464; Phys. Rev. D {\bf 58}, 116002 (1998), hep-ph/9809521; S. Coleman and S. L. Glashow, Phys. Rev. D {\bf 59}, 116008 (1999), hep-ph/9812418.

\bibitem{sme_calc} 
V. A. Kosteleck\'{y} and Ch. Lane, J. Math. Phys. {\bf 40} 6245 (1999), hep-ph/9909542; Phys. Rev. D {\bf 60}, 116010 (1999), hep-ph/9908504; R. Bluhm et. al, Phys. Rev. D {\bf 68}, 125008 (2003), hep-ph/0306190.

    \bibitem{sme_reviews}
    R. Bluhm, {\it Overview of the Standard Model Extension: Implications and Phenomenology of Lorentz Violation}, hep-ph/0506054; D. Mattingly, Living Rev. Rel. {\bf 8}, 5 (2005), gr-qc/0502097.

\bibitem{Andrianov_Giacconi_Soldati}
A. A. Andrianov, P. Giacconi, and R. Soldati, Grav. Cosmol. Suppl. {\bf 8N1}, 41 (2002), astro-ph/0111350.

    \bibitem{sme_eff_particles}
    R. Bluhm, V. A. Kosteleck\'{y}, and N. Russell, Phys. Rev. D {\bf 57}, 3932 (1998), hep-ph/9809543; Phys. Rev. Lett. {\bf 82}, 2254 (1999), hep-ph/9810269.

\bibitem{sme_vmm_theory} 
V. A. Kostelecky, Ch. Lane, and A. Pickering, Phys. Rev. D {\bf 65}, 056006 (2002), hep-th/0111123.

\bibitem{sme_vmm_from_phot} 
Ch. D. Carone, M. Sher, and M. Vanderhaeghen, Phys. Rev. D {\bf 74}, 077901 (2006), hep-ph/0609150.

\bibitem{Zhukovsky_Lobanov_Murchikova}
V. Ch. Zhukovsky, A. E. Lobanov, and E. M. Murchikova, Phys. Rev. D {\bf 73}, 065016 (2006),
hep-ph/0510391.

\bibitem{Ebert_Zhukovsky_Razumovsky}
D. Ebert, V. Ch. Zhukovsky, and A. S. Razumovsky, Phys. Rev. D {\bf 70}, 025003 (2004),
hep-th/0401241.

    \bibitem{sme_eff_astro}
    T. Jacobson, S. Liberati, and D. Mattingly, Annals Phys. {\bf 321}, 150 (2006), astro-ph/0505267.

\bibitem{Altschul}
    B. Altschul, Phys. Rev. D {\bf 74}, 083003 (2006), hep-ph/0608332; Phys. Rev. Lett. {\bf 96}, 201101 (2006), hep-ph/0603138; Phys. Rev. D {\bf 72}, 085003 (2005), hep-th/0507258.

\bibitem{Montemayor_Urrutia}
R. Montemayor and L. F. Urrutia, Phys. Rev. D {\bf 72}, 045018 (2005), hep-ph/0505135.

\bibitem{Ellis_Mavromatos_Sakharov}
J. Ellis, N. E. Mavromatos, and A. S. Sakharov, Astropart. Phys. {\bf 20}, 669 (2004),
astro-ph/0308403.

\bibitem{Furry}
W. H. Furry, Phys. Rev. {\bf 81}, 115 (1951).

\bibitem{Schwinger_vmm}
J. Schwinger, Phys. Rev. {\bf 73}, 416 (1948).

\bibitem{Kostelecky_liv_grav} 
V. A. Kosteleck\'{y}, Phys. Rev. D {\bf 69}, 105009 (2004), hep-th/0312310.

\bibitem{Shapiro_liv_grav} 
I. L. Shapiro, Phys. Rept. {\bf 357}, 113 (2002), hep-th/0103093.

\bibitem{liv_from_chiral_fermions} 
G. E. Volovik, JETP Lett. {\bf 70}, 1 (1999), hep-th/9905008; G. E. Volovik and A. Vilenkin, Phys. Rev. D {\bf 62}, 025014 (2000), hep-ph/9905460.

\bibitem{Pauli_vmm}
W. Pauli, Rev. Mod. Phys. {\bf 13}, 203 (1941).

\bibitem{Ternov_Bagrov_Zhukovsky}
I. M. Ternov, V. G. Bagrov, and V. Ch. Zhukovsky, Vestnik Mosk. Univ., Fiz. Astron. {\bf 7}, No. 1, 30 (1966).

\bibitem{book_rel_el}
A. A. Sokolov and I. M. Ternov, {\it Synchrotron Radiation}, Akademie-Verlag, Berlin (1968); {\it Radiation from Relativistic Electrons}, American Institute of Physics, New York (1986).

\bibitem{book_hep_rad}
{\it Synchrotron Radiation Theory and its Development: in Memory of I.~M.~Ternov} (Series on synchrotron radiation techniques and application, Vol. 5) (Ed. V.~A.~Bordovitsyn), Singapore, World Scientific (1999).

\bibitem{Schwinger_sr} 
J. Schwinger, Proc. Nat. Acad. Sci. USA, {\bf 40}, 132 (1954).

\bibitem{book_qed} 
V. B. Berestetskii, E. M. Lifshitz, and L. P. Pitaevskii, {\it Course of Theoretical Physics}, Vol. 4: {\it Quantum Electrodynamics}, Pergamon Press, Oxford (1982).

\end{thebibliography}
\end{document}